\documentclass[aps,superscriptaddress,twocolumn,pra,longbibliography,floatfix]{revtex4-1}

\usepackage{amsmath,amsthm,amsfonts,amssymb,bm,graphicx,color,mathpazo,times, braket}
\usepackage[colorlinks={true}, citecolor={blue}, filecolor={blue}, linkcolor={blue}, urlcolor={blue}]{hyperref}
\usepackage[caption=false]{subfig}
\usepackage{graphicx}
\usepackage{graphicx}
\usepackage{soul}
\usepackage{tikz}
\usepackage{multirow}
\usetikzlibrary{positioning}

\usepackage{amsmath}
\usepackage[caption=false]{subfig}  

\allowdisplaybreaks
\begin{document}
\title{Optimizing quantum sensing networks via genetic algorithms and deep learning}
\author{Asghar Ullah}
\email{aullah21@ku.edu.tr}
\affiliation{Department of Physics, Ko\c{c} University, 34450 Sar\i yer, Istanbul, T\"urkiye}
\author{\"Ozg\"ur E. M\"ustecapl\i o\u glu}
\email{omustecap@ku.edu.tr}
	\affiliation{Department of Physics, Ko\c{c} University, 34450 Sar\i yer, Istanbul, T\"urkiye}
	\affiliation{T\"UBİTAK Research Institute for Fundamental Sciences (TBAE), 41470 Gebze, T\"urkiye}
\author{ Matteo G. A. Paris}
\email{matteo.paris@fisica.unimi.it}
\affiliation{Dipartimento di Fisica, Universita di Milano, I-20132 Milan, Italy}
\thanks{*Corresponding author: aullah21@ku.edu.tr}

\date{\today} 
\begin{abstract}
We investigate the optimization of graph topologies for quantum sensing networks designed to estimate weak magnetic fields. The sensors are modeled as spin systems governed by a transverse-field Ising Hamiltonian in thermal equilibrium at low temperatures. Using a genetic algorithm (GA), we evolve network topologies to maximize a perturbative spectral sensitivity measure, which serves as the fitness function for the GA. For the best-performing graphs, we compute the corresponding quantum Fisher information (QFI) to assess the ultimate bounds on estimation precision. To enable efficient scaling, we use the GA-generated data to train a deep neural network, allowing extrapolation to larger graph sizes where direct computation becomes prohibitive. Our results show that while both the fitness function and QFI initially increase with system size, the QFI exhibits a clear non-monotonic behavior---saturating and eventually declining beyond a critical graph size. This reflects the loss of superlinear scaling of the QFI, as the narrowing of the energy gap signals a crossover to classical scaling of the QFI with system size. The effect is reminiscent of the microeconomic law of diminishing returns: beyond an optimal graph size, further increases yield reduced sensing performance. This saturation and decline in precision are particularly pronounced under Kac scaling, where both the QFI and spin squeezing plateau or degrade with increasing system size. We also attribute observed even-odd oscillations in the spectral sensitivity measure and QFI to quantum interference effects in spin phase space, as confirmed by our phase-space analysis. These findings highlight the critical role of optimizing interaction topology---rather than simply increasing network size---and demonstrate the potential of hybrid evolutionary and learning-based approaches for designing high-performance quantum sensors.

\end{abstract}
\maketitle
\section{Introduction}\label{Sec:intro}

Quantum sensing leverages inherently quantum features—such as coherence, entanglement, and squeezing—to surpass classical limits in the precision measurement of physical parameters~\cite{RevModPhys.89.035002,Giovannetti2011,Giovannetti2011Science,PhysRevLett.96.010401,Giorda_2018,PhysRevLett.128.040502,PhysRevLett.113.103004, PhysRevResearch.5.043184,MONTENEGRO20251,mukhopadhyay2024current}. A prominent example is magnetic field sensing, where quantum systems such as qubits or spins act as probes to detect weak magnetic fields with high sensitivity~\cite{PhysRevX.5.031010,PhysRevA.99.062330,PhysRevLett.104.133601,PhysRevA.109.032608,PhysRevLett.120.260503,Albarelli_2017,PhysRevLett.129.120503,Danilin2018,PhysRevLett.116.240801,PhysRevLett.115.190801}. While quantum protocols can offer substantial gains, their performance is often hindered by decoherence and environmental noise under realistic conditions. Despite these challenges, significant progress has been made in practical implementations, particularly in atomic magnetometry~\cite{ PhysRevLett.104.093602,PhysRevLett.109.253605,PhysRevLett.111.143001,PhysRevLett.110.160802,PhysRevLett.113.103004,Lucivero2014}. These advances highlight the importance of optimizing both the structure and dynamics of quantum sensors to reduce noise-induced degradation. Most theoretical and experimental efforts have focused on simple geometries or regular interaction patterns, such as spin chains and lattices~\cite{PhysRevLett.121.020402,PhysRevLett.124.120504,PhysRevA.99.062330,PhysRevB.103.L220407}. However, recent studies suggest that structured or networked interaction topologies may offer new pathways to enhance metrological performance~\cite{PhysRevLett.120.080501,PhysRevE.104.014136,Abiuso_2024}.

Network-based models offer a natural framework for describing interactions in quantum systems, where qubits are represented by nodes and their couplings by edges. Such graph-based descriptions have proven useful in exploring a range of quantum phenomena, including quantum transport~\cite{Chisholm_2021,Kurt_2023,PhysRevE.106.024118,ausilio2025memory,Annoni2024}, coherence preservation~\cite{PhysRevA.109.012424,Elster2015}, and sensing~\cite{PhysRevE.104.014136,Claudia,ullah2025configuration,campbell2025inferring}. In the context of quantum metrology~\cite{PhysRevLett.96.010401, Giovannetti2011}, the quantum Fisher information (QFI)~\cite{Paris2009} provides a fundamental bound on the precision with which a parameter can be estimated. Because QFI is highly sensitive to the underlying interaction topology, identifying optimal graph configurations for enhanced sensing is a challenging task, made difficult by the combinatorial growth of possible topologies. While many-body systems have been extensively studied for sensing applications~\cite{MONTENEGRO20251}, the specific role of graph structure in shaping metrological performance remains insufficiently explored. At the same time, emerging quantum technologies demand flexible strategies for engineering and optimizing interaction networks~\cite{Gebhart2023, Ban_2021}, further motivating the study of structured spin networks for improving sensitivity to weak magnetic fields.

In this work, we explore the optimization of graph-structured quantum sensors 
\cite{palmieri2021multiclass,palmieri2024enhancing,gianani2023multiparameter}
for magnetic field estimation using a genetic algorithm (GA)~\cite{Goldberg1988,KONAK2006992,eiben2005evolutionarycomputing}. The spin networks are governed by a transverse-field Ising Hamiltonian and are assumed to be in thermal equilibrium with a low-temperature bath. To quantify the sensitivity of each graph configuration, we employ two complementary metrics: the spectral sensitivity measure $D_n$, which captures the perturbative response of the energy spectrum~\cite{ullah2025configuration}, and the QFI, which establishes the ultimate precision bound for parameter estimation~\cite{Paris2009}.

To identify optimal topologies, we employ a GA to evolve graph structures by maximizing the fitness function $D_n$, beginning with an initial population of connected, path-like graphs. At each generation, the algorithm applies standard evolutionary operations—selection, crossover, and mutation. For the best-performing graphs identified via the spectral sensitivity measure \( D_n \), we compute the corresponding QFI. Among these, an important reference case is given by fully connected (complete) graphs, in which every spin interacts with all the others~\cite{wilson1979introduction}. To ensure meaningful thermodynamic behavior in fully-connected spin systems, we apply Kac scaling~\cite{Kac1963}, which rescales the interaction strength by a factor of $1/N$. This normalization ensures that the energy per spin remains finite as the system size increases, thereby preventing divergences in the thermodynamic limit. Spin squeezing, introduced by Kitagawa and Ueda~\cite{PhysRevA.47.5138}, which is also connected to quantum metrology~\cite{Tóth_2014}, provides a criterion for multipartite entanglement that is directly useful for enhanced precision measurements. In this context, the rescaled QFI $F_Q/N$ serves as a generalized spin-squeezing witness~\cite{RevModPhys.90.035005}, and is used throughout this work.
Our optimization results reveal two key phenomena: (i) a \textit{diminishing returns effect}, where superlinear scaling of the QFI saturates or even declines with increasing system size; and (ii) \textit{even-odd oscillations} in both QFI and spin squeezing, particularly evident in the absence of Kac scaling. To verify and explain these behaviors, we analyze complete graphs as representative topologies that exhibit strong quantum correlations and Dicke-like entanglement. In the Kac-scaled regime, the QFI saturates and spin squeezing decreases with system size, confirming the loss of superlinear scaling in the thermodynamic limit, where the system becomes extensive and effectively classical. By contrast, in the non-Kac-scaled regime, superlinear scaling persists even at larger $N$, with the QFI exhibiting superlinear growth in some cases. Furthermore, we find that the even-odd oscillations in QFI and spin squeezing originate from quantum interference effects in spin phase space, rather than from simple spectral parity. These findings underscore that increasing system size alone is insufficient to enhance metrological performance. Instead, careful optimization of interaction scaling and network topology is essential.

To extend our analysis to larger systems---where direct optimization becomes computationally intractable---we train a deep neural network (DNN) using GA-generated data for both even and odd values of \( N \). The DNN takes either \( D_n \) or the QFI as input and predicts both quantities for unseen system sizes, enabling efficient extrapolation. This hybrid strategy circumvents the need for full quantum simulations of large graphs, providing a scalable surrogate model for performance estimation.
Notably, our results also show that the genetic algorithm converges rapidly---often within just a few generations---further enhancing computational efficiency. These findings support the use of \( D_n \) as a reliable proxy for QFI, significantly lowering the cost of designing optimized quantum sensing networks.

The remainder of the paper is organized as follows. In Sec.~\ref{Sec:methods}, we describe the methods employed in this study, including the genetic algorithm, the spectral sensitivity measure $D_n$, and the QFI. In Sec.~\ref{Sec:Results}, we present the outcomes of the graph optimization process for different parameters, investigate quantum features and finite-size effects, and employ a deep neural network to extrapolate the results to larger graph sizes.. Finally, Sec.~\ref{conclusion} provides a summary of our main findings.

\section{Methods}\label{Sec:methods}
\subsection{Model description}
We consider a system of interacting spins or qubits described by a transverse-field Ising Hamiltonian defined on path-like connected graphs. Each vertex of the graph represents a qubit, and the edges define pairwise interactions between them. 
The Hamiltonian reads
\begin{equation}\label{model}
H = -J_{\text{eff}} \sum_{(i,j) \in E} \sigma_z^{(i)} \sigma_z^{(j)} - h \sum_{i=1}^N \sigma_x^{(i)},
\end{equation}
where $J_{\text{eff}}$ is the effective coupling strength (see below) between connected spins, $E$ denotes the set of edges of the graph, and $h$ is the strength of the transverse magnetic field. The operators $\sigma_\alpha^{(i)}$ (where $\alpha=x,y,z$) are Pauli matrices acting on the $i$-th qubit. 

We distinguish between two choices for the effective coupling $J_{\text{eff}}$:
In the unscaled (bare) model, we set $J_{\text{eff}} = J/2$.
In the Kac-scaled model, used when discussing thermodynamic behavior, we set $J_{\text{eff}} = J/2N$ to ensure an extensive total interaction energy in the large-$N$ limit. Throughout the paper, we restrict to the antiferromagnetic case with $J=-1$, unless stated otherwise.

We assume that the system is coupled to a low-temperature bath and is at thermal equilibrium. The Gibbs thermal state of the system is given by
\begin{equation}\label{GTS}
    \rho_T = \frac{e^{-\beta H}}{\mathcal{Z}},
\end{equation}
where $\beta=1/T$ (we set $k_B=1$) denotes the inverse temperature and the partition function $\mathcal{Z}$ is given by
\begin{equation}
    \mathcal{Z} = \mathrm{Tr}\left(e^{-\beta H}\right) = \sum_n e^{-\beta E_n},
\end{equation}
 where \(\{E_n, |\psi_n\rangle\}\) are the eigenvalues and eigenstates of the Hamiltonian \(H\), i.e., \(H|\psi_n\rangle = E_n |\psi_n\rangle\).

\subsection{The genetic algorithm}\label{Sec:GeneticAlgorithm}
\begin{figure}[h!]
    \centering
    \includegraphics[scale=0.6]{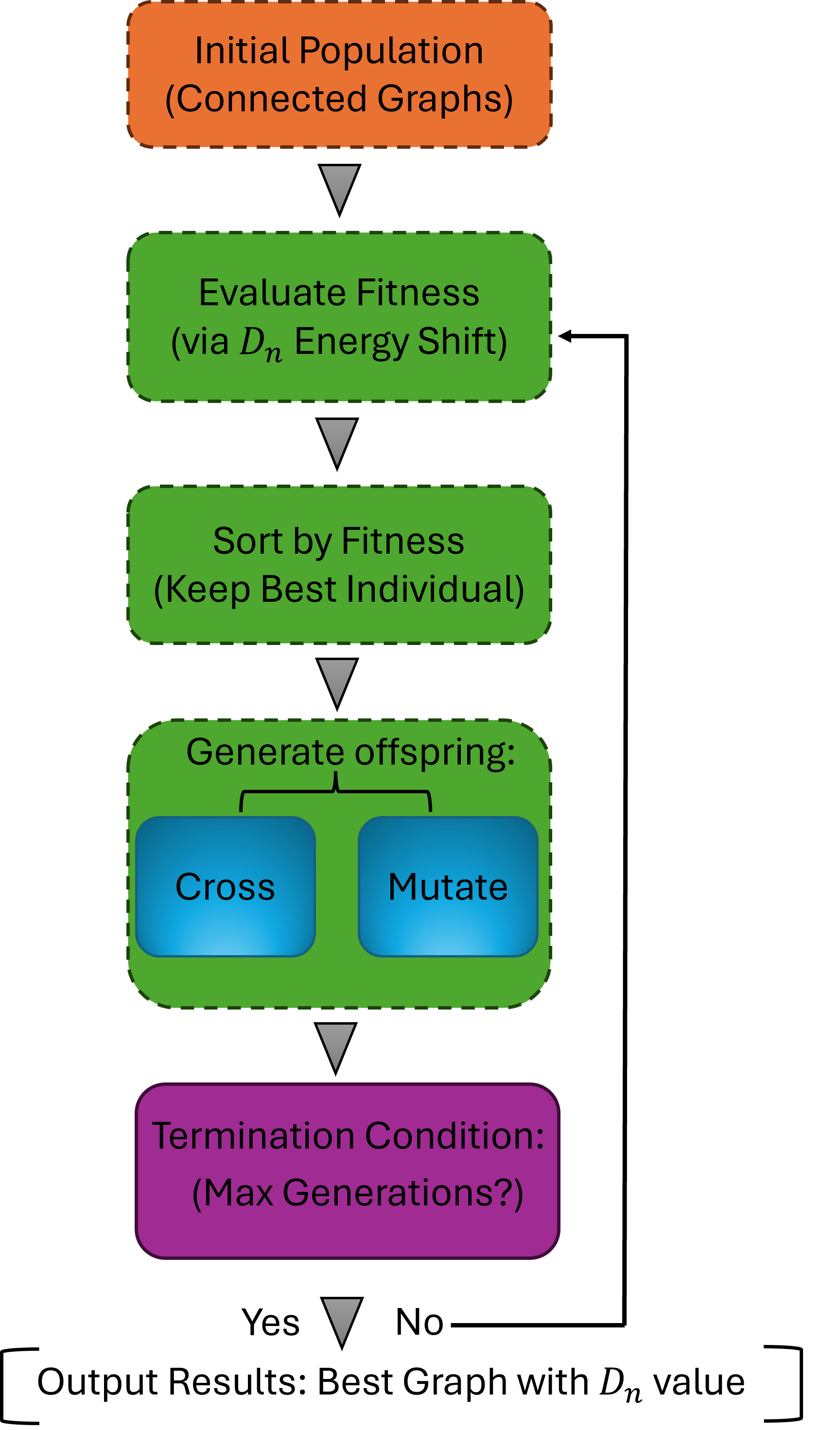}
    \caption{Genetic algorithm workflow.}
    \label{fig:GA}
\end{figure}

To explore optimal graph topologies for enhancing quantum sensitivity, we employed a GA tailored for discrete combinatorial optimization over graph topologies. Each individual in the population represents a connected graph with a fixed number of vertices \( N \), encoded by its adjacency structure.
A schematic overview of the genetic algorithm structure is shown in Fig.~\ref{fig:GA}. 
The GA proceeds through the following standard evolutionary steps:
 
 \textbf{Initialization:} An initial population $p$ of connected graphs is generated. For $N=1$, there is only one possible connected graph, which is a single node with no edges. While for  \( N > 1 \), a path graph is used as a seed structure, and random edges are added to introduce diversity.
    
 \textbf{Fitness Evaluation:} Each candidate graph is evaluated using a fitness function based on the spectral sensitivity measure \( D_n \). This quantity quantifies the perturbative sensitivity of the low-lying energy spectrum of the system under a small transverse magnetic field  $h$ (see Sec.~\ref{Dn} for more details).

 \textbf{Selection:} A subset of high-performing graphs is chosen based on their fitness scores. These selected graphs serve as parents for producing the next generation.
    
 \textbf{Crossover:} Pairs of parent graphs are combined by merging their edges into a child graph. To maintain diversity, with a 50\% probability, an additional edge is randomly added between two previously unconnected nodes in the child graph.
    
 \textbf{Mutation:} To introduce stochastic variation, random edges are added to the graph with a predefined mutation probability, without removing existing edges. Specifically, a single new edge is randomly added between unconnected node pairs if any exist. This "add-only" mutation helps explore new graph topologies while preserving existing connectivity. For reproduction, the top 50\% of graphs by fitness are
selected for reproduction via deterministic truncation to maintain the solution
quality.

\begin{figure}[t!]
    \centering
    \includegraphics[scale=0.36]{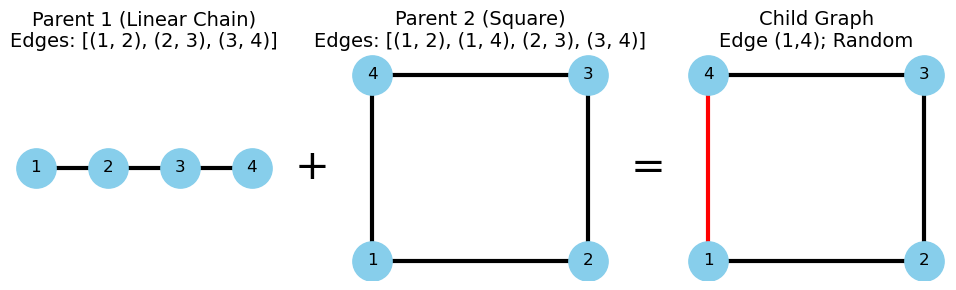}
    \caption{Example of graph evolution through genetic operations for \( N = 4 \) nodes. 
\textbf{Left to right:} 
(1) \textit{Parent 1} is a linear (path) graph with edges \((1,2), (2,3), (3,4)\). 
(2) \textit{Parent 2} is a square graph with edges \((1,2), (2,3), (3,4),(1,4)\). 
(3) \textit{Child (Crossover)} combines common edges from both parents, and the \textit{Mutated Child} further modifies the child by adding another random edge (1,4) (red colored), increasing the graph's connectivity.}
    \label{fig:CM}
\end{figure}

To visualize the combined effect of crossover and mutation, Fig.~\ref{fig:CM} shows an illustrative example for $N=4$. Two parent graphs—a linear chain and a square graph—are combined via crossover, resulting in a child graph that merges their common edges. Mutation then adds a new random edge to the child, increasing its connectivity. This example illustrates how diversity and exploration are introduced during GA evolution.
    
 \textbf{Elitism:} The best-performing graph from each generation is preserved unaltered into the next generation to ensure the current optimal solution is retained throughout the evolution.

This process is iterated over a fixed number of generations $n_G$. At each generation, the best graph with its edges and its corresponding \( D_n \) value is recorded to monitor progress. Rather than aiming to explore behavior at asymptotically large system sizes, this evolutionary approach focuses on efficiently identifying optimal graph topologies that maximize magnetic field sensitivity.

\begin{figure*}[t!]
    \centering
    \includegraphics[scale=0.75]{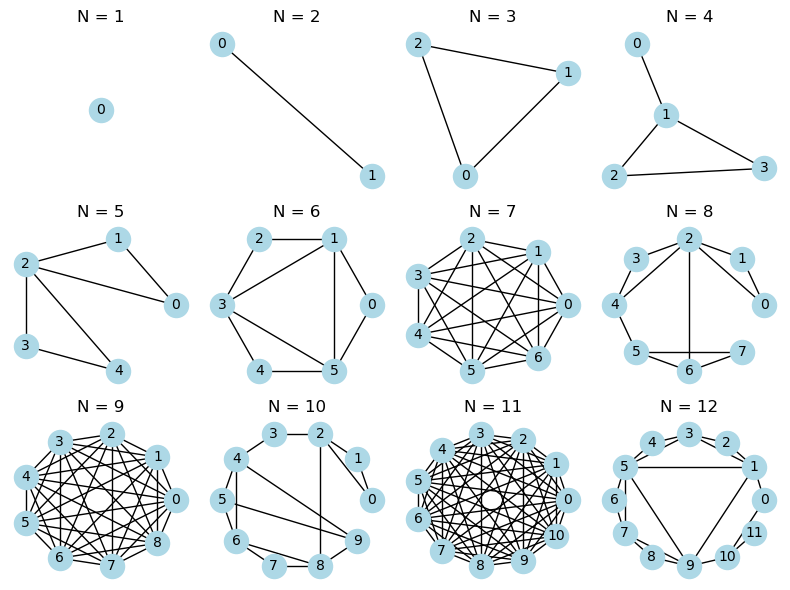}
    \caption{Optimal connected graph structures obtained via the genetic algorithm for \( N = 1 \) to \( N = 12 \). Each graph with the maximum $D_n$ value maximizes the QFI under the given model, reflecting the topologies most sensitive to magnetic field estimation. These graphs are obtained for the parameters $T=0.08$, $h=0.05$, $p=100$, and $n_G=15$.}
    \label{fig:graphs}
\end{figure*}
\subsection{Fitness function $D_n$}\label{Dn}

To quantify the sensitivity of a graph-structured quantum system to an external perturbation $h$, we employ the spectral sensitivity measure \(D_n\) as the fitness function for the GA. While the QFI is a natural candidate for evaluating the sensitivity of a quantum system and could, in principle, be used as a fitness function in our graph optimization scheme, its computation is significantly more expensive, particularly for large system sizes and dense graphs. To validate our choice, we compared both approaches exactly for small systems and found that optimization with either the QFI or $D_n$ yields the same optimal graph structures and sensitivity. This consistency arises because $D_n$, defined as the Euclidean distance between the ground states at $h=0$ and $h>0$, is maximized precisely when the system exhibits enhanced response to perturbations, which directly correlates with a larger QFI. Therefore, in practice, we employ $D_n$ as an efficient and scalable proxy for QFI in our optimization process.

Given a graph defined by its edge set, we construct the transverse-field Ising Hamiltonian at two values of the transverse field strength: the unperturbed case \(h=0\) and a small perturbation \(h > 0\) with the Hamiltonians such as \(\hat{H}(0)\) and \(\hat{H}(h)\), respectively.

We compute the lowest \(n\) eigenvalues of each Hamiltonian with $E_i(0)$ and $E_i(h)$, where $i=1, \ldots, n$. We define the fitness function as the Euclidean distance between the lowest $n$ eigenvalues (we set $n=2$ in our case), such as~\cite{ullah2025configuration}
\begin{equation}
D_n(h) = \sqrt{\sum_{i=1}^n \bigl( E_i(h) - E_i(0) \bigr)^2}.
\end{equation}

This measure captures the shift in the low-energy manifold induced by the magnetic field perturbation, providing a scalar fitness score that guides the evolutionary search towards graphs exhibiting the largest spectral sensitivity.
\subsection{Quantum Fisher information for thermal states}\label{Sec:QFI}

We use QFI to assess the metrological performance of graph sensors for magnetic field estimation. To calculate the QFI for the thermal state of the graph-based quantum sensing networks, we follow the formalism developed in Ref.~\cite{PhysRevA.89.032128}. Given the Gibbs thermal state in Eq.~\eqref{GTS}, we define the Hermitian operator
\begin{equation}
G := -\beta H(h) - \ln \mathcal{Z},
\end{equation}
so that \( \rho = e^{G} \). Let \( G = \sum_j g_j |e_j\rangle \langle e_j| \) be the spectral decomposition of the operator \( G \), with orthonormal eigenvectors \( \{ |e_j\rangle \} \). Note that $g_j=-\beta\lambda_j-\ln{\mathcal{Z}}$, where $\lambda_j$ are the energy eigenvalues of $H$, i.e, $H|e_j\rangle=\lambda_j|e_j\rangle$.

To evaluate the QFI with respect to \( h \), we introduce the symmetric logarithmic derivative (SLD) \( L \), defined via
\begin{equation}
\frac{d\rho}{dh} = \frac{1}{2} (L \rho + \rho L).
\end{equation}
In the eigenbasis of \( G \), the matrix elements of \( L \) are given by
\begin{equation}
L_{jk} = f(g_j, g_k) \, \dot{G}_{jk}, \qquad \text{with} \quad \dot{G}_{jk} := \langle e_j | \dot{G} | e_k \rangle,
\end{equation}
where
\begin{equation}
f(g_j, g_k) =
\begin{cases}
1, & j = k, \\[4pt]
\displaystyle \frac{\tanh\left(\frac{g_j - g_k}{2}\right)}{\frac{g_j - g_k}{2}}, & j \neq k.
\end{cases}
\end{equation}
\begin{figure}[t!]
    \centering
    \subfloat[]{
    \includegraphics[scale=0.56]{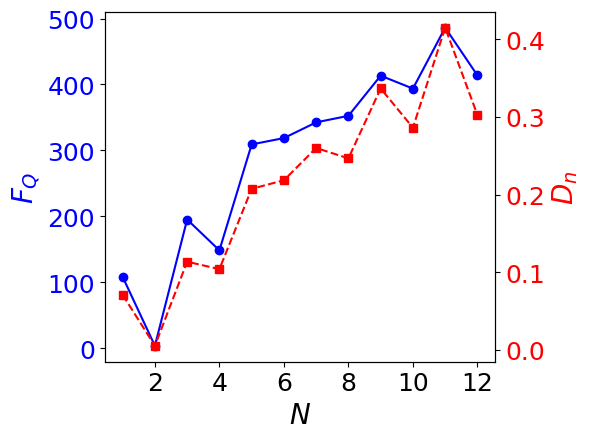}}\\
      \subfloat[]{
       \includegraphics[scale=0.56]{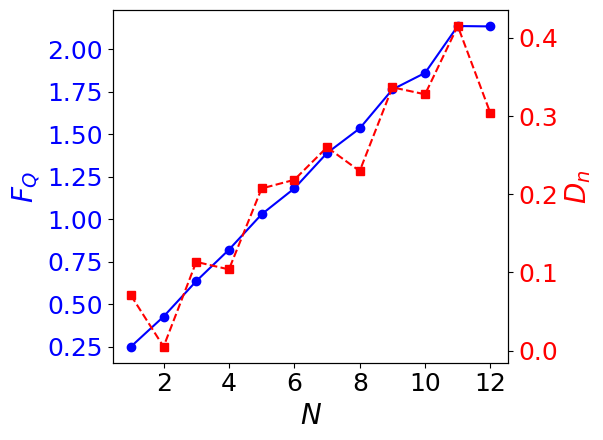}}
    \caption{QFI (blue) and $D_n$ (red) as a function of number of vertices $N$ for the optimal graphs shown in Fig.~\ref{fig:graphs} and using two different values of temperature (a) $T=0.08$ and (b) $T=2$ when the magnetic field is set to $h=0.05$. The rest of the parameters are fixed at $p=100$ and $n_G=15$.}
    \label{fig:DifferentT}
\end{figure}

Let \( X := \frac{dH}{dh} \). Then,
\begin{equation}
\dot{G} = -\beta X - \frac{d}{dh} \ln Z \cdot \mathbb{I},
\end{equation}
where $\mathbb{I}$ denotes the identity matrix. By defining $p_j := e^{g_j}=e^{-\beta\lambda_j}/\mathcal{Z}$, we obtain
\begin{equation}
\dot{G}_{jk} =
\begin{cases}
-\beta \left( X_{jj} - \sum_m p_m X_{mm} \right), & j = k, \\[4pt]
-\beta X_{jk}, & j \neq k,
\end{cases}
\end{equation}
with \( X_{jk} = \langle e_j | X | e_k \rangle \). The QFI is then given by
\begin{equation}
F_Q = \mathrm{Tr}(\rho L^2) = \sum_{j,k} p_j |L_{jk}|^2 = \sum_{j,k} p_j \left| f(g_j, g_k) \,\dot{G}_{jk} \right|^2.
\end{equation}
This expression provides an efficient route to computing the QFI in thermal equilibrium, based on the spectral decomposition of \( G \) and the matrix elements of the derivative operator \( X \).
\begin{figure}[t!]
    \centering
    \includegraphics[scale=0.55]{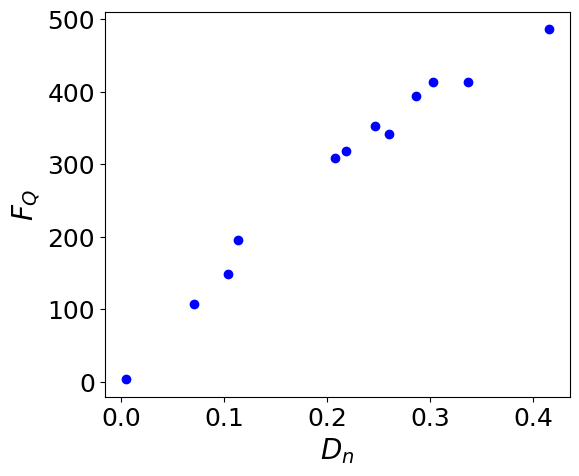}
    \caption{QFI as a function of the fitness function $D_n$ for the optimal graphs shown in Fig.~\ref{fig:graphs}. The parameters are set to $T=0.08$, $h=0.05$, and $N=12$. The rest of the parameters are the same as in Fig.~\ref{fig:DifferentT}.}
    \label{fig:QFI-vs-Dn}
\end{figure}
\section{Results}\label{Sec:Results}
In Sec.~\ref{GA-Results}, we present the results obtained from the genetic algorithm and analyze how different parameter sets affect the performance of graph-based quantum sensors. In Sec.~\ref {quantum_features}, we discuss quantum features and scalings in both non-Kac (Sec.~\ref{Section_IIIA}) and Kac scaling (Sec.~\ref{Section_IIIB}) regimes. Finally, in Sec.~\ref{DNN} we use the GA-generated data to train a deep neural network, enabling extrapolation of sensitivity trends to larger graph sizes.
\subsection{Genetic algorithm-based optimization of $D_n$ and QFI}\label{GA-Results}

We now present the results from the GA optimization of graph topologies aimed at maximizing magnetic field sensitivity via the spectral sensitivity measure \(D_n\). Figure~\ref{fig:graphs} shows the optimal graphs obtained at \(T=0.08\) and \(h=0.05\), corresponding to the highest \(D_n\) and thus the maximum QFI. For completeness, we also examine how the optimal graph structures vary with the magnetic field strength \(h\), as detailed in Appendix~\ref{effect_h}. This analysis reveals how graph connectivity adapts during optimization to enhance sensitivity under different conditions.

To assess robustness, we verified that the GA, run with multiple random seeds and different initial populations, consistently converged to the same optimal topology (up to graph isomorphism) for the system sizes considered. We note that other gradient-free optimizers (e.g., particle swarm or simulated annealing) could, in principle, identify different candidate graphs~\cite{PhysRevResearch.4.043057}. However, establishing that an alternative topology is superior requires direct benchmarking using the same figures of merit (QFI and $D_n$). Since exhaustive QFI evaluation across the full graph space is feasible only for small $N$, we rely on exact enumeration for these sizes and use the GA as an efficient heuristic for larger systems. Consequently, while other algorithms could in principle be applied, $D_n$ provides a physically motivated and efficiently implementable objective for the GA, and its heuristic, non-analytic structure can make it nontrivial to integrate into alternative optimization schemes.

We begin by presenting the results for the optimal values of $D_n$ and the QFI as functions of the system size $N$ for two different temperature regimes, as shown in Fig.~\ref{fig:DifferentT}. The transverse magnetic field is fixed in the weak-field regime, specifically at $h = 0.05$. 

In the low-temperature case ($T = 0.08$), Fig.~\ref{fig:DifferentT}(a) clearly shows that both $D_n$ and QFI exhibit oscillations between even and odd values of $N$, which are characteristic of finite-size quantum effects. The quantum interference between spin states defined along different quantization axes gives rise to these even–odd oscillations observed in the QFI as a function of system size. These oscillations reflect alternating constructive and destructive interference in the collective spin phase space. The underlying mechanism is analogous to the interference phenomena first discussed by W.~Schleich and co-workers in the context of spin and atomic coherent states~\cite{Schleich1987,PhysRevA.38.1177} (see also Refs.~\cite{PhysRevA.48.1854, PhysRevE.97.042127}), where interference between distinct spin orientations produces oscillatory quantum signatures. Our results, therefore, represent the same fundamental physics manifested in the behavior of the QFI. A similar phenomenon has been reported in chains of coupled critical resonators~\cite{Alushi2025}, where the observed oscillations were likewise linked to the structure of the system’s phase space.

In addition, the QFI values are significantly higher in the low-temperature regime, exhibiting both even and odd oscillations. In contrast, in the high-temperature regime ($T = 2$), Fig.~\ref{fig:DifferentT}(b) shows that the QFI is substantially suppressed at higher temperatures, highlighting the detrimental impact of thermal noise on magnetic field estimation. The QFI exhibits an approximately linear scaling with \(N\), consistent with the SQL~\cite{Giovannetti2004}. These results indicate that even a small finite temperature, such as $T = 0.08$, is sufficient to reduce the quantum interference effects that enhance sensitivity. As temperature increases, thermal fluctuations progressively wash out these quantum features, suppressing the QFI and driving its behavior toward that of classical-like systems. Figure~\ref{fig:DifferentT} thus illustrates the detrimental impact of temperature on quantum features and metrological performance. We discuss these quantum signatures and finite-size behaviors in more detail in Sec.~\ref{quantum_features}. To further quantify the effectiveness of using $D_n$ as the GA fitness function, Fig.~\ref{fig:QFI-vs-Dn} shows the explicit correlation between $D_n$ and the resulting QFI for the optimal graphs. The data in Fig.~\ref{fig:QFI-vs-Dn} shows that larger values of $D_n$ consistently yield higher QFI, demonstrating the effectiveness of $D_n$ as a fitness criterion.
\begin{figure}[t!]
    \centering
    \includegraphics[scale=0.56]{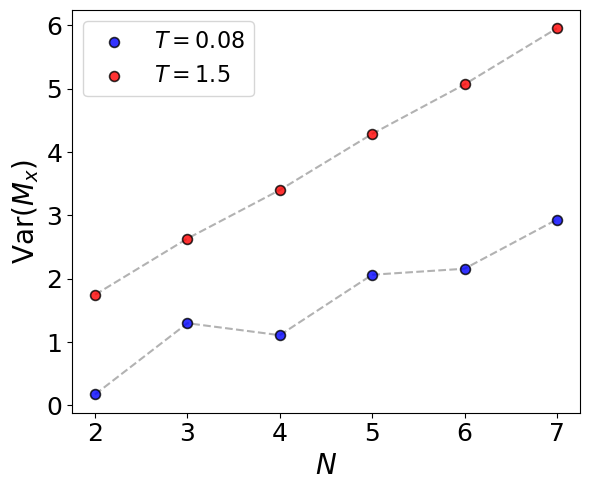}
    \caption{Variance of magnetization ($\text{Var}(M_x)$) as a function of number of spins $N$ for complete graphs when $T=0.08$ (blue) and $T=1.5$ (red). The magnetic field is fixed at $h=0.05$, and we consider the antiferromagnetic case $J=-1$. The results are obtained for connected graphs under the transverse field Ising Hamiltonian.}
    \label{fig:magnetization}
\end{figure}
\subsection{Quantum features and finite size effects}\label{quantum_features}
\subsubsection{Magnetization, $\chi_x$, and basic QFI scaling}\label{Section_IIIA}
In this section, we discuss the diminishing returns in the superlinear scaling of the QFI, the even-odd oscillations in $N$, and we analyze the energy gap and magnetization variance.

The total transverse magnetization operator is defined as
\begin{equation}
    M_x = \sum_{i=1}^N \sigma_i^x,
\end{equation}
where \(\sigma_i^x\) is the Pauli \(X\) operator acting on site \(i\). The magnetic susceptibility \(\chi_x\) quantifies the response of the system’s magnetization to the transverse field \(h\), defined as
\begin{equation}
    \chi_x = \frac{\partial \langle M_x \rangle}{\partial h}.
\end{equation}
By the fluctuation-dissipation theorem, the variance of the total magnetization relates to the susceptibility and temperature as
\begin{equation}
    \mathrm{Var}(M_x) = \langle M_x^2 \rangle - \langle M_x \rangle^2 = N k_B T \chi_x = \frac{N}{\beta} \chi_x.
\end{equation}
We calculate $\chi_x$ using the partition function, as follows
\begin{equation}
    \chi_x=\frac{1}{N\beta}\frac{\partial^2}{\partial h^2}\ln\mathcal{Z}.
\end{equation}
This relation allows us to evaluate quantum fluctuations and magnetic response from susceptibility calculations. Figure~\ref{fig:magnetization} illustrates the variance of magnetization $M_x$ as a function of system size $N$ for distinct connected graphs, examined in both low and high temperature limits. At low temperatures ($T=0.08$), the variance of $M_x$, denoted as $\text{Var}(M_x)$, exhibits even-odd oscillations with respect to $N$, which gradually decrease in amplitude as $N$ increases. In contrast, at high temperatures ($T=1.5$), these oscillations disappear, and $\text{Var}(M_x)$ becomes linearly proportional to $N$.
\begin{figure}[t!]
    \centering
    \includegraphics[scale=0.55]{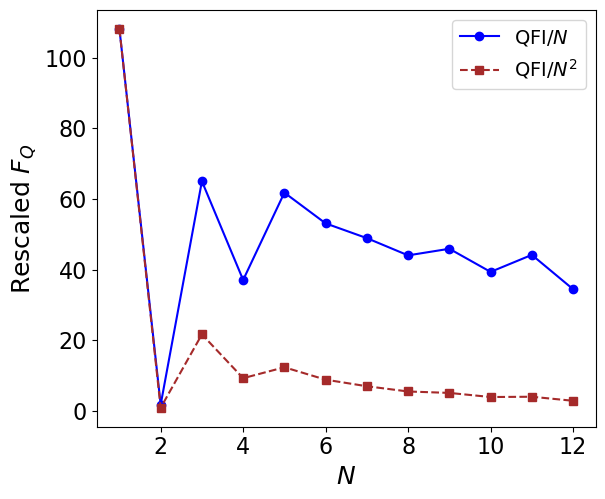}
    \caption{Rescaled QFI as a function of system size $N$ for the optimal graphs shown in Fig.~\ref{fig:graphs}. The plot shows \(F_Q / N\) (blue solid line)  as a function of system size \(N\), indicating the gain per spin and \(F_Q / N^2\) (brown dashed line)  versus \(N\), revealing the scaling behavior and the onset of diminishing returns as $N$ increases. The rest of the parameters are the same as in Fig.~\ref{fig:magnetization}.
}
    \label{fig:QFIN}
\end{figure}
\begin{figure}[t!]
    \centering
    \subfloat[]{
    \includegraphics[scale=0.56]{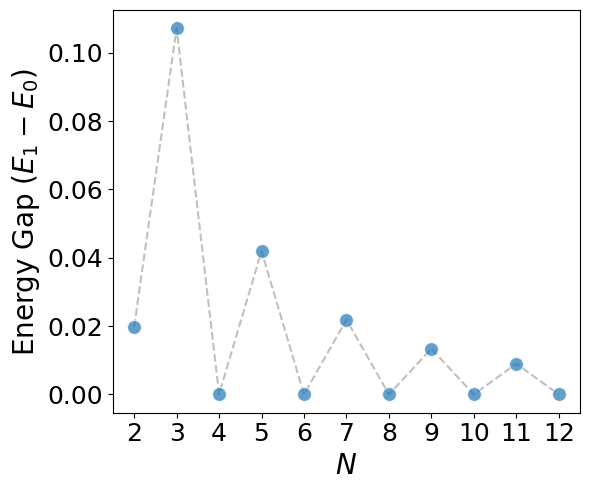}}\\
    \subfloat[]{\includegraphics[scale=0.56]{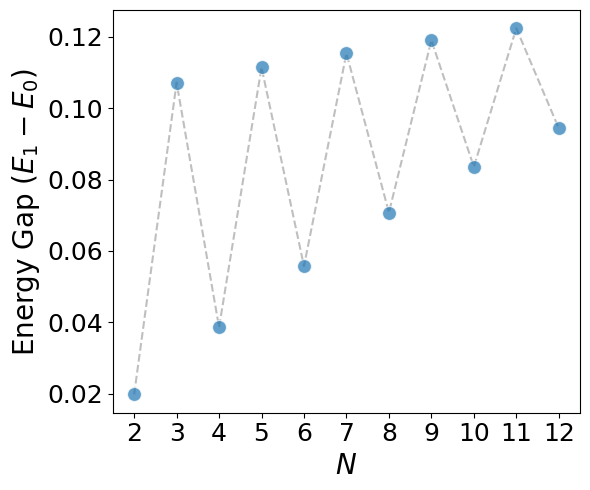}}
    \caption{Energy gap $E_1-E_0$ between the ground and first excited states as a function of system size $N$ for (a) a 1D transverse-field Ising linear chain with periodic boundary conditions and (b) fully connected (complete) graphs. The parameters are set to $J=-1$ and $h=0.1$.}
    \label{fig:energy gap}
\end{figure}
It is worth noting that the variance of magnetization $M_x$, rather than the susceptibility, is proportional to the QFI~\cite{PhysRevX.8.021022}. This relationship explains why fluctuations of $M_x$ scale proportionally with the system size at high temperatures. Consequently, the observed $\text{Var}(M_x)$ aligns with the behavior of QFI: at $T=0.08$ and $h=0.05$, it accurately reproduces QFI predictions (see Fig.~\ref{fig:DifferentT}), and at $T=1.5$ and $h=0.05$, it demonstrates the anticipated linear dependence on $N$, consistent with QFI behavior.

Regarding the observed diminishing returns, we attribute this to finite-size quantum effects. We clarify that "diminishing returns" here means increasing \(N\) yields only linear, not nonlinear, improvements in precision. Initially, the QFI exhibits nonlinear scaling with the number of spins, $N$, which constitutes a genuine quantum signature. However, this behavior is not indefinitely robust: as $N$ increases, the influence of quantum interference diminishes, leading to a crossover where the QFI grows linearly with $N$, akin to classical scaling. This transition is expected as in the thermodynamic limit, where quantum effects are progressively washed out by thermal fluctuations or averaging over large system sizes.
\begin{figure*}[t!]
    \centering
    \includegraphics[scale=0.42]{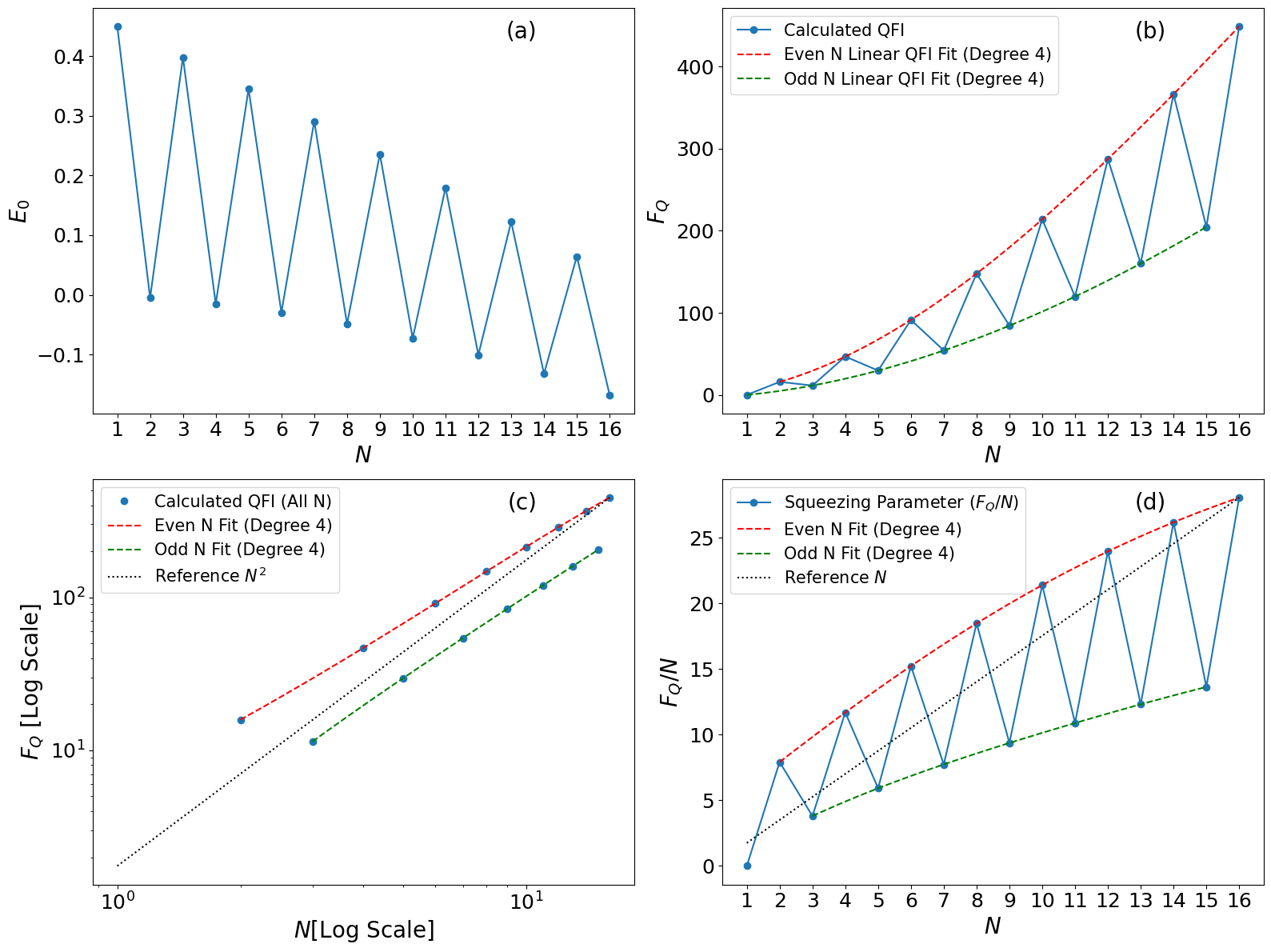}
    \caption{Results for complete graphs without Kac scaling (non-Kac regime) at $T=0$.  
(a) Ground state energy \(E_0\) as a function of system size \(N\) (non-Kac, complete graphs).  
(b) QFI \(F_Q\) as a function of \(N\) on a linear scale (non-Kac, complete graphs).  
(c) QFI \(F_Q\) as a function of \(N\) in log-log scale with separate fourth-order polynomial fits for even and odd \(N\) (non-Kac, complete graphs).  
(d) Spin squeezing parameter \(F_Q / N\) (Kitagawa-Ueda definition, as a measure of multipartite entanglement) versus \(N\), with fourth-order polynomial fits for even and odd \(N\) (non-Kac, complete graphs). Here, we set \(J = -1\) and \(h = 0.05\).
}\label{fig:No_Kac}
\end{figure*}
To quantify this crossover from quantum to classical behavior and to clarify the onset of diminishing returns in QFI scaling, we compute the \textit{energy gap} $\Delta E = E_1 - E_0$ between the ground and first excited states. The energy gap provides a natural scale against which external perturbations or thermal energies $T$ can be compared. 
A finite energy gap isolates the ground state from excitations, preserving coherence and quantum correlations that support enhanced QFI scaling. 
When the gap closes, low-energy modes mix, coherence is reduced, and the system approaches classical behavior. 
Thus, the gap behavior helps reveal the spectral origin of the even–odd oscillations. Analyzing the behavior of the energy gap as a function of system size thus provides insight into the spectral origin of the observed even--odd oscillations. In addition, we relate this choice to the spectral sensitivity measure $D_n$, which quantifies the energy shift between ground states for $h>0$ and $h=0$~\cite{ullah2025configuration}. Since $D_n$ effectively captures the system’s response to external perturbations, its dependence on the energy gap provides further motivation for using the gap as a natural indicator of the quantum-to-classical crossover. In particular, as the gap diminishes with increasing system size, quantum features are progressively washed out, consistent with the expected crossover from nonlinear to linear QFI scaling. When $T \gg \Delta E$ or when $N$ is sufficiently large, the energy gap effectively closes, leading to diminishing returns in metrological quantities such as the QFI. To illustrate the scaling behavior explicitly, in Fig.~\ref{fig:QFIN}, we plot the scaled quantities \(F_Q / N\) and \(F_Q / N^2\) as functions of system size \(N\) for the optimal graphs shown in Fig.~\ref{fig:graphs}. After strong fluctuations at small \(N\), the quantity \(F_Q/N\) settles to a slowly decreasing trend for \(N \ge 3\), indicating that $F_Q/N$ does not grow further with system size. In contrast, \(F_Q/N^2\) rapidly drops from its initial value and remains close to zero for larger \(N\), confirming the onset of diminishing returns, indicating a transition from quantum-enhanced to classical-like scaling.
\begin{figure*}[t!]
    \centering
    \includegraphics[scale=0.42]{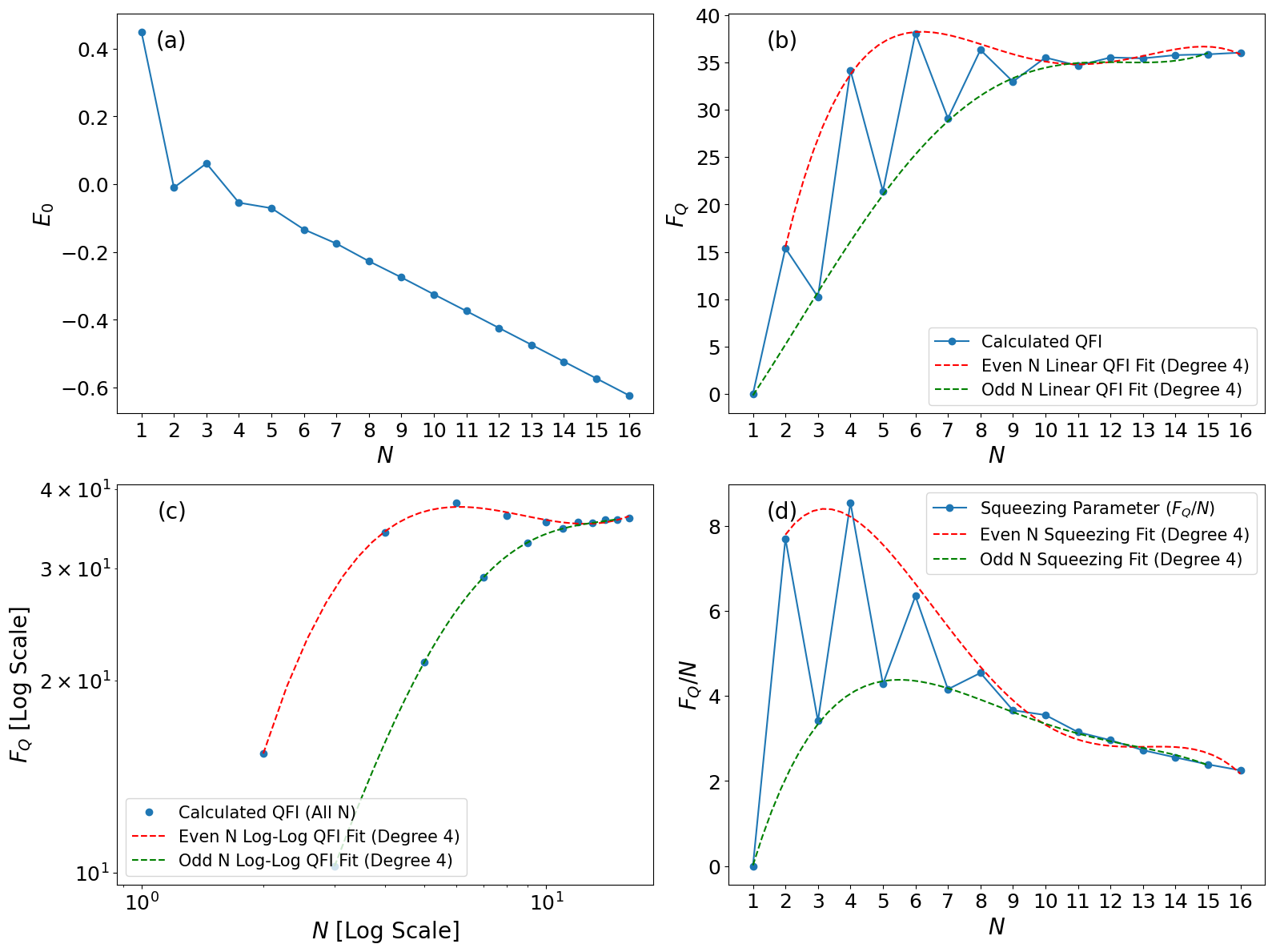}
    \caption{Results for complete graphs with Kac scaling (Kac regime) at \(T=0\).  
(a) Ground state energy \(E_0\) as a function of system size \(N\) (Kac, complete graphs).
(b) QFI \(F_Q\) as a function of \(N\) on a linear scale (Kac, complete graphs).  
(c) QFI \(F_Q\) as a function of \(N\) in log-log scale with separate fourth-order polynomial fits for even and odd \(N\) (Kac, complete graphs).  
(d) Spin squeezing parameter \(F_Q / N\) (Kitagawa-Ueda definition, as a witness of multipartite entanglement) versus \(N\), with fourth-order polynomial fits for even and odd \(N\) (Kac, complete graphs). Here, we set \(J = -1\) and \(h = 0.05\).}
    \label{fig:with_Kac_scaling}
\end{figure*}

In Fig.~\ref{fig:energy gap}, we compute the energy gap of the $1D$ transverse field Ising model with periodic boundary conditions, defined on both linear chain graphs and complete graphs. As shown in Fig.~\ref{fig:energy gap}(a), the linear chain rapidly loses quantum features as the system size $N$ increases, with the energy gap closing quickly. In contrast, complete graphs preserve quantum characteristics over a much broader range of $N$, as shown in Fig.~\ref{fig:energy gap}(b), although they too eventually lose superlinear scaling of the QFI. Interestingly, as $N$ increases, the optimal graph structures tend to resemble complete graphs, implying that the observation of quantum-to-classical crossover might require even larger $N$ in such highly connected systems.

Therefore, while features such as enhanced QFI and nontrivial magnetization persist at small $N$, these effects gradually weaken as the system size increases. Our simulations indicate that the transition to diminishing returns—manifested as the onset of linear QFI scaling—is not readily apparent for small systems (e.g., $N \leq 12$), particularly in highly connected graphs like the complete graphs shown in Fig.~\ref{fig:graphs}. Nevertheless, the progressive closing of the energy gap with increasing $N$ signals that enhanced quantum sensitivity will inevitably vanish, consistent with expectations in the thermodynamic limit.

\subsubsection{Finite-size scaling, spin squeezing, and the role of graph structure at $T=0$}\label{Section_IIIB}
In this section, we investigate the finite-size scaling behavior of key physical quantities relevant to our model at $T=0$: the QFI $F_Q$, the ground state energy $E_g$, and the spin squeezing parameter $F_Q/N$. We consider two ways of scaling the interaction strength. In the Kac-scaled case, the coupling is normalized by the system size, giving an effective coupling
$J_{\mathrm{eff}} = J/2N$.
This scaling ensures that the total interaction energy remains extensive (i.e., proportional to \(N\)) in the thermodynamic limit~\cite{CAMPA200957}.
In contrast, the non-Kac-scaled case retains the bare coupling
$J_{\mathrm{eff}} = J/2$, which does not depend on \(N\). These two approaches allow us to compare the effects of system-size-dependent and system-size-independent interaction strengths.

\begin{table*}[t]
\centering

\caption{Comparison of key behaviors in the non-Kac-scaled and Kac-scaled regimes in Figs.~\ref{fig:with_Kac_scaling} and \ref{fig:No_Kac}. 
All results correspond to the antiferromagnetic case ($J=-1$) and zero temperature ($T=0$) regime.}
\begin{tabular}{p{3.9cm} p{6.1cm} p{6.1cm}}
\hline\hline
\textbf{Feature} & \textbf{Non–Kac scaling} ($J_{\mathrm{eff}} = J/2$) 
& \textbf{Kac scaling} ($J_{\mathrm{eff}} = J/2N$) \\[3pt]
\hline
Interaction type & unnormalized couplings; superextensive total energy 
& Normalized couplings; extensive total energy \\[4pt]
Energy gap $\Delta E$ & Remains finite for larger $N$; slower closing
& Rapidly closes with $N$, leading to classical limit \\[4pt]
QFI scaling with $N$ & Superlinear, $F_Q \!\sim\! N^{\alpha>1}$;
persistent quantum advantage
& Saturates or decreases; crossover to linear $F_Q\!\propto\! N$ (SQL) \\[4pt]
Spin squeezing $F_Q/N$ & Remains high; oscillations between even/odd $N$ persist
& Decreases beyond $N\!\approx\!10$; oscillations suppressed with $N$ \\[4pt]
 Graph topology & Dense or complete graphs remain optimal
& Same topology but reduced metrological gain \\[4pt]
Physical interpretation & Non-extensive collective correlations enhance sensitivity;
quantum regime
& Extensive energy scaling suppresses interference and squeezing;
classical-like regime \\[4pt]
\hline\hline
\end{tabular}
\label{tab:kaccomparison}

\end{table*}


We compute the ground state energy $E_0$ of Hamiltonian $H$, and evaluate the QFI specifically for estimating a small perturbation in the transverse field, given by~\cite{PhysRevX.8.021022}:
\begin{equation}\label{QFI:M_x}
F_Q = 4 \left( \langle M_x^2 \rangle - \langle M_x \rangle^2 \right),
\end{equation}
where $M_x=\sum_{i=1}^N\sigma^x_i$ is the collective spin operator along the x-direction. This expression~\eqref{QFI:M_x} is valid only for the ground state. Finally, the metrologically relevant spin-squeezing parameter is defined as the normalized QFI~\cite{Tóth_2014}
\begin{equation}
\xi^2 = \frac{F_Q}{N}.
\end{equation}
Figures~\ref{fig:No_Kac} and~\ref{fig:with_Kac_scaling} present a comparative analysis of finite-size scaling in complete graphs without and with Kac scaling, respectively, at zero temperature. To analyze the finite-size scaling behavior of the QFI and spin squeezing, we perform polynomial fits of degree up to $4$ on the data as a function of the system size $N$, given by
\begin{equation}
    y(N)= \sum_{k=0}^d a_kN^k,\quad\quad d\le4
\end{equation}
where $y(N)$ represents the quantity of interest (e.g., $F_Q$ or $F_Q/N$, $a_k$ are the coefficients of the polynomial, and the degree $d$ is chosen based on the number of data points to avoid overfitting. 
While higher-than-quadratic scaling is not physically expected for qubit systems in the thermodynamic limit, the fourth-order polynomial fit is used here only to capture finite-size effects that arise for small $N$. When Kac scaling is not applied, these finite-size corrections can lead to apparent superlinear trends, whereas with Kac scaling, the asymptotic scaling remains linear for large $N$. Thus, the higher-order fit serves as a numerical interpolation to describe the crossover from small-$N$ quantum-interference behavior to the linear extensive regime. In addition, for the QFI $F_Q$, we also perform polynomial fits in the logarithmic scale:
\begin{equation}
    \log y(N)=\sum_{k=0}^db_k(\mathrm{Log}N)^k,
\end{equation}
and to test whether QFI follows a scaling law of the form $F_Q\sim N^{\alpha}$.

First, we observe that the law of diminishing returns manifests only in the Kac-scaled case (Fig.~\ref{fig:with_Kac_scaling}), where both the QFI and the spin squeezing parameter $F_Q /N$ saturate or even decrease with increasing $N$, indicating the loss of superlinear scaling in the thermodynamic limit. In contrast, in the non-Kac case (Fig.~\ref{fig:No_Kac}), both the QFI and spin squeezing continue to increase with $N$, demonstrating persistent scaling of both QFI and spin squeezing with system size. Second, while the ground state energy $E_0$ exhibits diminishing even-odd oscillations with increasing $N$ in both cases, these oscillations persist in the QFI data, especially pronounced in the Kac-scaled model (Fig.~\ref{fig:with_Kac_scaling}). This indicates that the oscillations cannot be fully explained by energy spectrum features or parity effects alone, pointing instead to more subtle quantum origins that we explore further in the following section through phase-space analysis in Section~\ref{Phase_Space}.

\begin{figure*}[]
    \centering
    \includegraphics[scale=0.53]{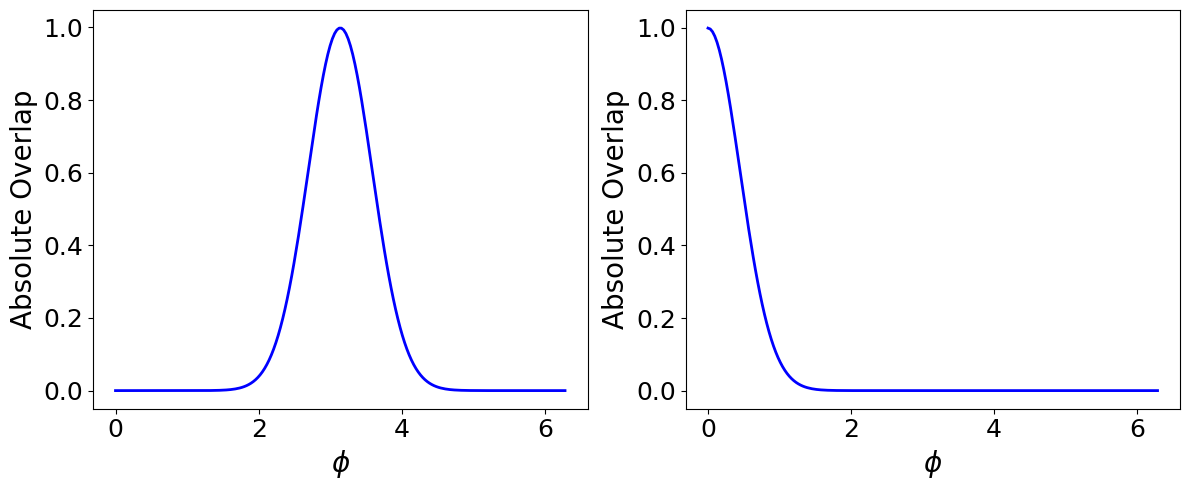}\\
    \includegraphics[scale=0.47]{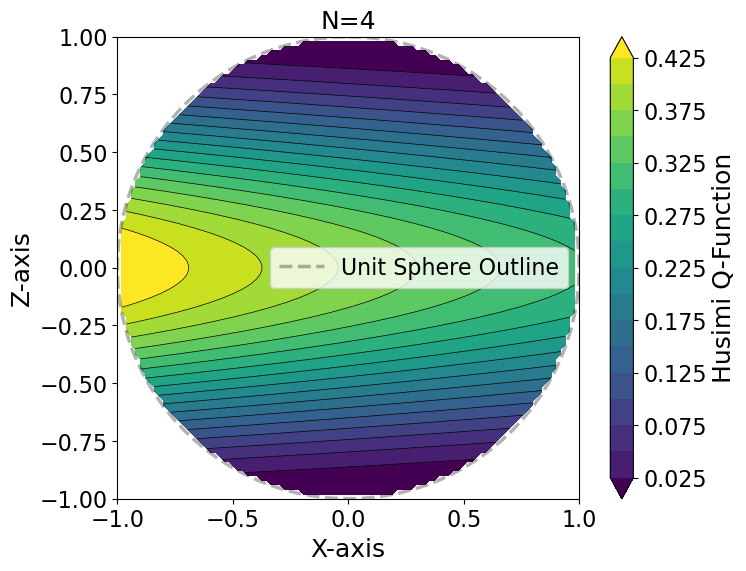}
    \includegraphics[scale=0.47]{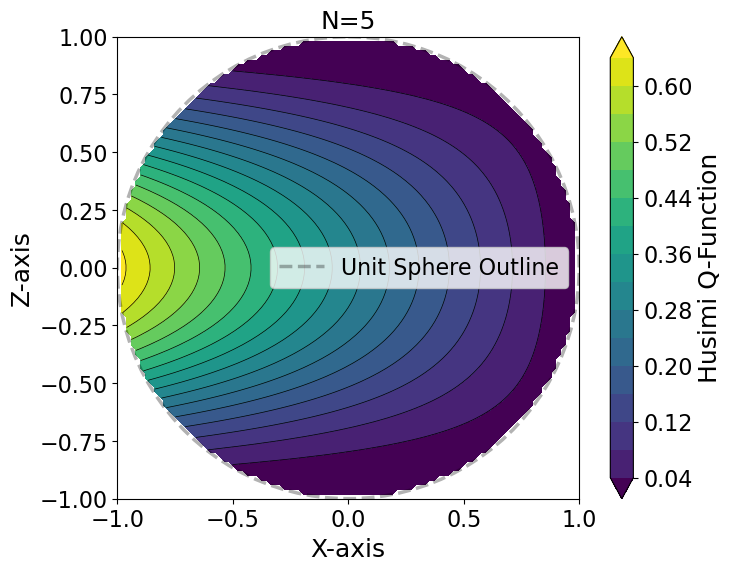}
    \caption{Top panel: Absolute overlap $|\langle \theta,\phi | \psi_{\mathrm{0}} \rangle|$ of the ground state $\ket{\psi_{\mathrm{0}}}$ with spin coherent states $\ket{\theta,\phi}$ as a function of the azimuthal angle $\phi$ along the equator ($\theta = \pi/2$) of the Bloch sphere for $N=4$ and $N=5$ spins. This plot illustrates the projection distribution of the ground state on the coherent spin basis. Bottom panel: Husimi Q-function contour plots of the ground states for \(N=4\) and \(N=5\) spins projected onto the ZX plane. The calculations are performed on complete graphs using Hamiltonian in Eq.~\eqref{model}. The rest of the parameters are set to $J=-1$ and $ h=0.05$.}
    \label{fig:Q_Functions}
\end{figure*}

The complete graph topology naturally supports Dicke-like collective states characterized by strong global entanglement. This is evident in the behavior of the spin squeezing parameter (Figs.~\ref{fig:No_Kac}(d) and \ref{fig:with_Kac_scaling}(d)), which serves as a witness for multipartite entanglement~\cite{Tóth_2014}. However, this witness does not fully quantify entanglement depth in large-spin systems. Furthermore, the persistence of squeezing-related oscillations in the non-Kac case correlates with phase-space interference effects, as evidenced by the Husimi function analysis we present in Section~\ref{Phase_Space}. This supports previous findings that spin squeezing is associated with interference structures in phase space, even without definitive multipartite entanglement characterization~\cite{PhysRevE.97.042127}.
Notably, in the non-Kac case, spin squeezing oscillations persist without attenuation, indicating robust multipartite entanglement as the system size grows. Conversely, in the Kac-scaled case, the squeezing parameter decreases steadily after $N\sim 10$, coinciding with a transition to linear QFI scaling, and exhibits the suppression of even-odd oscillations. The Kitagawa-Ueda squeezing parameter, computed perpendicular to the mean spin, indicates entanglement but is not a definitive multipartite measure for many-spin systems. Furthermore, since similar squeezing behavior occurs across multiple graph topologies for given parameters, thus, QFI scaling arises not solely from entanglement, but also from the topology-dependent spectral properties of the system. In addition to quantum correlations, spectral sensitivity~\cite{ullah2025configuration} plays a significant role in determining which graph topology achieves the highest QFI.

These findings highlight that the observed oscillations in QFI are not only a manifestation of quantized energy levels, emblematic of the first quantum revolution, but also a hallmark of the second quantum revolution, characterized by non-local correlations and quantum interference. In the next section, we elucidate these interference effects through a detailed phase-space analysis using the Husimi $Q$-function. We remark that temperature plays a secondary but predictable role: it tends to wash out these interference-induced oscillations and suppress squeezing, thereby reducing the benefits of non-Kac scaling. Nonetheless, even at low but finite $T$, the non-Kac case retains superlinear scaling of the QFI, although this requires the challenging engineering of equal-strength all-to-all couplings across many qubits, similar to the challenges encountered in scaling quantum processors with dense connectivity~\cite{RevModPhys.93.025001, Crawford2023, Künne2024}. To further clarify the contrasting scaling behaviors observed in the two regimes, we include Table~\ref{tab:kaccomparison}, which summarizes the main differences in scaling behaviors between the Kac and non-Kac regimes.

These insights underscore the need for optimizing graph structures tailored to realistic, finite-temperature conditions. While the complete graph is optimal at $T=0$, alternative graph topologies may offer enhanced robustness against thermal noise, preserving useful quantum correlations for quantum metrology. Identifying such topologies remains an open direction for both theoretical exploration and experimental realization.
\subsubsection{Phase space interference at $T=0$}\label{Phase_Space}
We analyze the origin of the even-odd oscillations observed in the QFI for optimal (complete) graphs, where each spin interacts equally with all others. These oscillations arise from phase-space interference between squeezed spin states and the eigenstates of the \(S_x\) operator, analogous to interference effects between photonic squeezed states and Fock states in quantum optical phase space \((x,p)\), which exhibit similar even-odd oscillations in photon number distributions~\cite{PhysRevE.97.042127}. Our analysis focuses on the zero-temperature regime (\(T=0\)), where such quantum interference effects are most pronounced.
\begin{figure}[t!]
    \centering
    \includegraphics[scale=0.72]{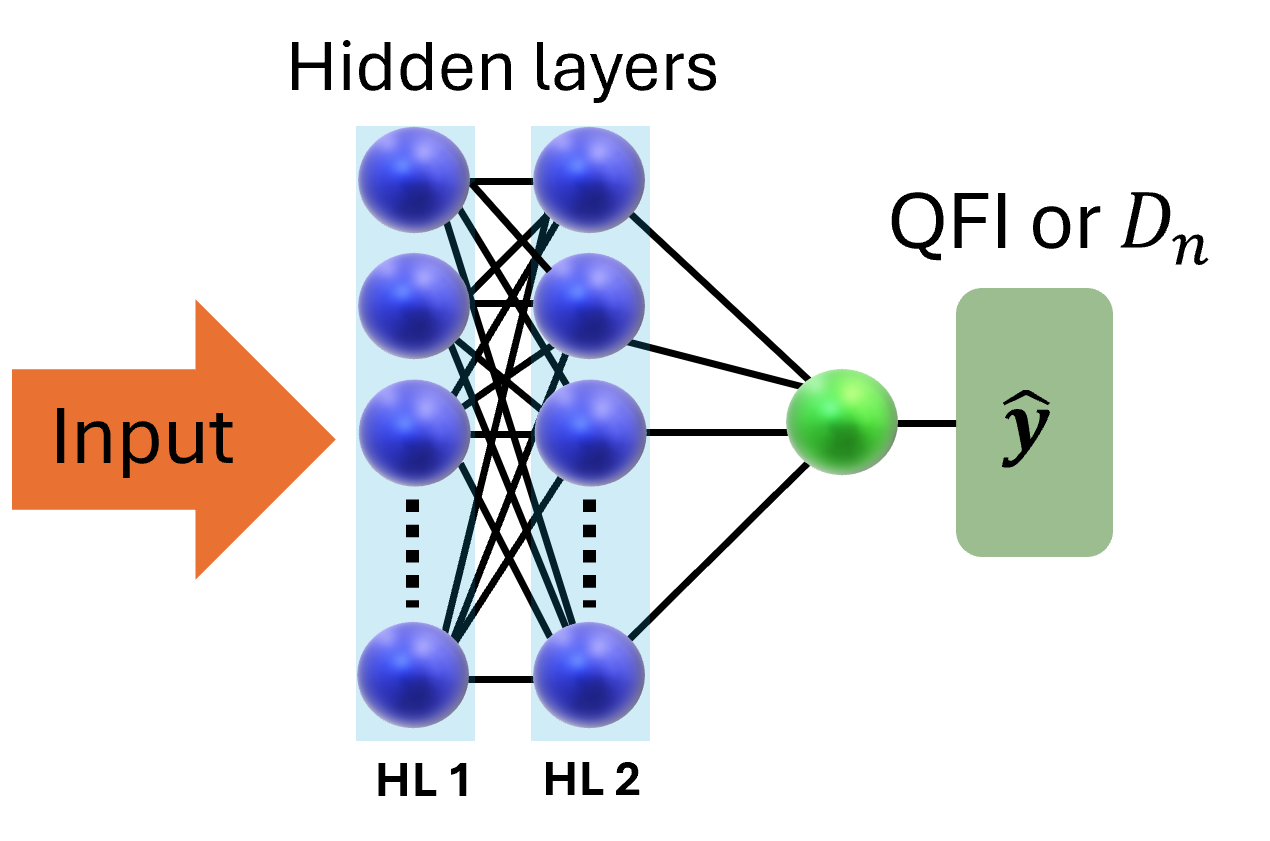}
    \caption{Architecture of the fully connected feedforward neural network used to predict $D_n$ or the QFI data obtained from GA. The input is a scalar number $N\in\mathbb{R}$ which represents the QFI of $D_n$ values. The network consists of two hidden layers with $64$ and $32$ neurons, respectively, each followed by a ReLU activation function. The final output produces a single scalar value, which corresponds to the predicted $D_n$ or the QFI values.}
    \label{fig:Nnetwork}
\end{figure}
We compute the Husimi Q-function as follows
\begin{equation}
    Q(\theta,\phi)=\frac{1}{\pi}|\bra{\theta,\phi}\ket{\psi_0}|^2,
\end{equation}
where $\ket{\psi_0}$ is the ground state of $H$ and $\ket{\theta,\phi}$ is the spin coherent state defined by
\begin{equation}
    \ket{\theta,\phi}=\bigotimes_{k=1}^N\big(\cos{(\frac{\theta}{2})\ket{0}_k}+e^{i\phi}\sin{(\frac{\theta}{2})\ket{1}_k}\big).
\end{equation}
We also compute the absolute overlap \( |\langle \theta = \pi/2, \phi | \psi_0 \rangle| \) as a function of the azimuthal angle \( \phi \), i.e., along the equator of the Bloch sphere (\( \theta = \pi/2 \)), as shown in Fig.~\ref{fig:Q_Functions} (top panel).
This provides a direct measure of how the ground state aligns with coherent states localized in the equatorial plane. Our results show that for \(N=4\), the overlap peaks sharply at \( \phi = \pi \), indicating that the ground state points predominantly along the negative X-axis. On the other hand, when \(N=5\), the peak shifts to \( \phi = 0 \), corresponding to the positive X-axis. Thus, the preferred orientation of the ground state alternates between even and odd \(N\), reflecting an even-odd parity effect in the phase space localization of the state.
\begin{figure}[t!]
    \centering
    \subfloat[]{
    \includegraphics[scale=0.55]{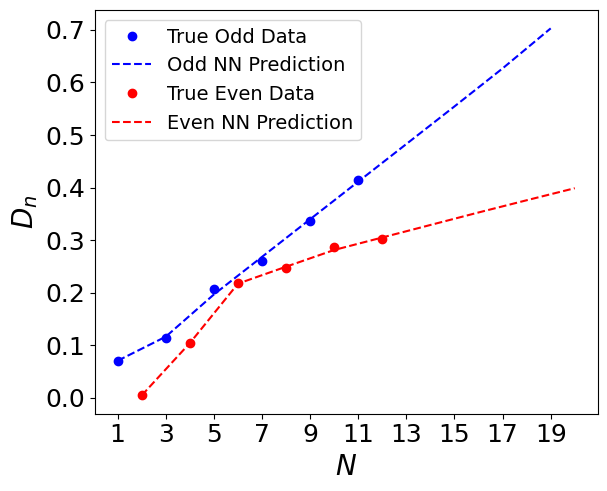}}\\
    \subfloat[]{
    \includegraphics[scale=0.55]{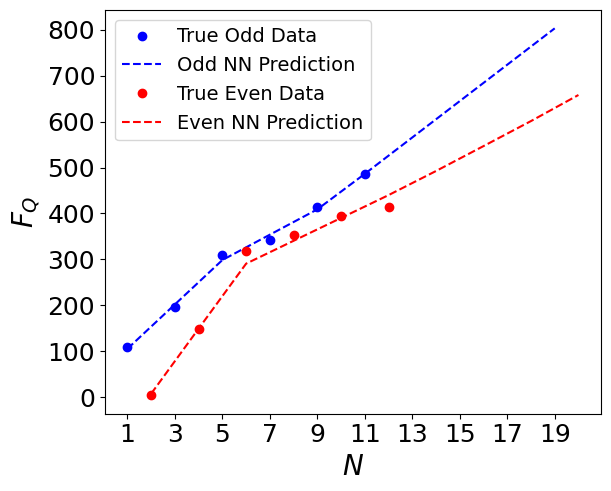}}
    \caption{Comparison of true and neural network predicted values of the perturbative sensitivity metric \(D_n\) (panel a) and QFI (panel b) for graphs with sizes \(N=1\) to \(12\) (for optimal graphs in Fig.~\ref{fig:graphs}). The model is trained on the known odd $N$ (blue circles) and even $N$ (red circles) data and accurately captures the underlying trends, enabling reliable extrapolation to larger graphs (\(N=13\) to \(21\), green squares). The parameters are set to $T=0.08$. $h=0.05$, $p=100$, $n_G=15$, and $\eta=0.001$.}
    \label{fig:NN}
\end{figure}
To further understand these oscillations, we calculate the mean spin \( \langle S_x \rangle \) via the Husimi integral:
\begin{equation}
    \langle S_x\rangle = \int_0^\pi \int_0^{2\pi} Q(\theta,\phi) \sin\theta \cos\phi \, d\phi \, d\theta.
\end{equation}
We observe oscillations in the variance of \(S_x\), which are mirrored in the behavior of the QFI. Notably, only the mean \( \langle S_x \rangle \) oscillates while  \( \langle S_x^2 \rangle \) does not. To visualize this, we plot the Husimi \(Q\)-function of the ground state and overlay it with the classical spin projection \( S_x = S \cos\phi \sin\theta \), where \(S = N/2\) in Fig.~\ref{fig:Q_Functions}(bottom panel). These contour plots of Husimi Q functions reveal that for even \(N\), the dominant Husimi contours center around the origin, while for odd \(N\), they are biased away from it. This alternating localization pattern in phase space directly contributes to the observed even-odd oscillations in the QFI. Therefore, these even-odd oscillations in QFI and variance are direct manifestations of the underlying phase space interference in the ground state, as visualized via the Husimi Q-function and overlap analysis.

\subsection{Learning graph sensitivity metrics with a deep feedforward neural network}\label{DNN}
 To better understand the trend of QFI versus $N$, we employ separate deep neural networks (DNNs) trained on odd and even $N$ subsets (see Fig.~\ref{fig:Nnetwork}) to predict the perturbative sensitivity \(D_n\) and QFI for graph sizes beyond those optimized by the GA. In particular, we utilize separate deep neural networks (DNNs) trained on odd and even $N$ data subsets to capture their distinct behaviors.

The input to each network is the scalar spin number $N\in\mathbb{R}$, and the output is the predicted $D_n$ or QFI $F_Q$ values. Formally, the networks approximate functions, such as
\begin{equation}
    \mathcal{F}_\theta^{\text{odd}} : \mathbb{R} \to \mathbb{R}, \quad N \mapsto \hat{F}_Q^{\text{odd}}(N) \quad \text{or} \quad \hat{D}_n^{\text{odd}}(N),
\end{equation}
and 
\begin{equation}
    \mathcal{F}_\theta^{\text{even}} : \mathbb{R} \to \mathbb{R}, \quad N \mapsto \hat{F}_Q^{\text{even}}(N) \quad \text{or} \quad \hat{D}_n^{\text{even}}(N),
\end{equation}
where \(\theta\) denotes the network parameters. Given training data \(\{(N_i, F_{Q,i})\}_{i=1}^M\) or \(\{(N_i, D_{n,i})\}_{i=1}^M\) separated into odd and even subsets, the networks minimize the mean squared error (MSE) loss, such that
\begin{equation}
\mathcal{L}(\theta) = \frac{1}{M} \sum_{i=1}^M \big( \mathcal{F}_\theta(N_i) - \hat{y}_i \big)^2,
\end{equation}
where \(y_i\) denotes the target value ( either \(F_{Q,i}\) or \(D_{n,i}\)) and $\hat{y}_i=\mathcal{F}_\theta(N_i)$ is the prediction. The deep neural network models, illustrated in Fig.~\ref{fig:Nnetwork}, are trained separately on odd and even system sizes \(N\). Each network consists of two fully connected hidden layers with 64 and 32 neurons, respectively. The hidden layers use the rectified linear unit (ReLU) activation function. Training is performed by minimizing the MSE loss function using the Adam optimizer~\cite{kingma2017adam}, with the entire dataset used as a batch for each update. The models are trained for $4000$ epochs with a fixed learning rate of \(\eta=0.001\).

The DNN is trained separately on even and odd system sizes and generalizes to predict values \(\hat{y}\) for larger \(N\) beyond the training range. Figure~\ref{fig:NN} presents the results, showing excellent agreement between true and predicted values for \(D_n\) and QFI up to \(N=12\). Extrapolated predictions for \(N=13\) to \(N=21\) are shown as dashed lines. As illustrated in Fig. 13(a), the predicted $D_n$ for odd $N$ increases approximately linearly with system size, while for even $N$, the growth is slower. In fact, it is the difference $\Delta D_n = D^{(N+1)}_n - D^{(N)}_n$ which gradually decreases as $N$ increases. Figure~\ref{fig:NN}(b) compares the QFI predictions to true data, demonstrating that the model successfully captures the increasing trend of the QFI with N for both even and odd subsets and extrapolates accurately beyond the training range. The neural network also shows excellent agreement with training data and robust extrapolation to larger graphs. These results underscore the model’s ability to capture size-dependent trends and scaling behavior of \(D_n\) and QFI effectively.

\section{Conclusion}\label{conclusion}

We employed a computational framework that combines a GA with neural networks to optimize graph-based networks for quantum sensing of magnetic fields. We modeled the interactions within each graph by a transverse-field Ising Hamiltonian, assuming the system is in thermal equilibrium at a known temperature. We used two key metrics to assess each network topology, defined by its graph structure for its ability to estimate a weak magnetic field: the spectral sensitivity measure \( D_n \), used as the fitness function, and the QFI, which quantifies the metrological performance of the graph sensors. We employed the GA to explore the large combinatorial space of possible graph configurations, identifying those with enhanced sensitivity to external fields. We guided the optimization by the spectral sensitivity measure \( D_n \), which we showed to correlate strongly with the QFI while being computationally more efficient.

Our results demonstrate that the GA converges rapidly, often within the first 10 generations, and frequently identifies optimal graph configurations within the initial generations. The QFI exhibits a superlinear increase with system size, signaling enhanced quantum sensitivity. Despite this promising behavior, beyond a critical size \( N \), the QFI saturates and transitions to linear scaling with \( N \), reflecting the onset of the standard quantum limit where sensitivity improves only as $1/\sqrt{N}$. This behavior highlights a phenomenon of diminishing returns, which becomes apparent when Kac scaling is applied. Since Kac scaling normalizes the interaction strength and reveals extensive behavior. The transition towards classical behavior is associated with the closing of the energy gap. Specifically, for complete graphs, both the QFI and spin squeezing tend to saturate---or even decline---as \( N \) increases. Notably, this diminishing returns effect is challenging to observe for finite \( N \) without Kac scaling, which could lead to misleading interpretations of scaling trends. For example, in the high-temperature regime (\( T = 2 \)), the QFI scales linearly with \( N \), indicating the absence of superlinear scaling and thus no enhanced sensitivity as $N$ grows. Beyond the global scaling trends, we also observe even-odd oscillations in both $D_n$ and QFI, which we attribute to quantum interference effects. We explored these effects through a phase-space analysis based on the Husimi \( Q \)-function, highlighting the direct correspondence between interference effects and QFI oscillations. Our analysis reveals that the even-odd oscillations in the QFI are direct manifestations of phase interference patterns in the ground state, as visualized by the Husimi \( Q \)-functions. Therefore, phase-space interference emerges as the fundamental origin of the oscillations observed in our results~\cite{Schleich1987,PhysRevA.38.1177,PhysRevA.48.1854, PhysRevE.97.042127}. Furthermore, our findings indicate that only graph topologies with an even number of nodes (\( N \ge 6 \)) are sensitive to variations in the magnetic field.

To probe system-size scaling beyond the optimized datasets, we separately trained a deep neural network on the even and odd \( N \) data identified by the genetic algorithm. Using \( D_n \) and the QFI as inputs, the network efficiently learned the sensitivity metrics and successfully extrapolated them to larger system sizes. Our results reveal that, for odd \( N \), the QFI increases linearly with system size, whereas for even \( N \), it initially exhibits a decreasing trend before transitioning to linear scaling with \( N \).

Taken together, our results underscore the critical role of interaction topology, quantum interference, and scaling behavior in determining quantum sensing performance. We show that neural networks can serve as efficient surrogate models, enabling exploration of large-scale sensing architectures without the need for exhaustive quantum simulations. By integrating evolutionary algorithms with machine learning, our data-driven framework offers a scalable and resource-efficient approach to designing quantum sensors. These findings highlight the importance of topology-aware design strategies in advancing both the theoretical understanding and practical implementation of next-generation quantum sensing technologies.
\section*{Data availability statement}
The data that support the findings of this article are not
publicly available. The data are available from the authors
upon reasonable request.
\section*{Acknowledgments}
This work is supported by the Scientific and Technological Research Council (T\"UBİTAK) of T\"urkiye under Project Grant No. 123F150 and by EU and MUR under project BaC-PE00000002-QBETTER.
\appendix

\begin{figure}[t!]
    \centering
    \includegraphics[scale=0.55]{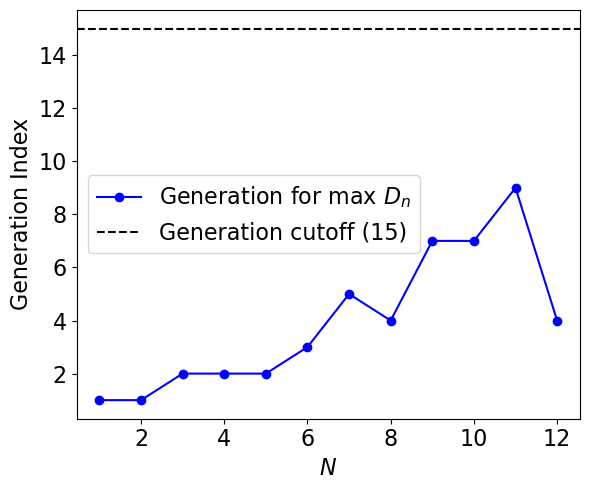}
    \caption{Generation index at which the maximum value of $D_n$ is found for each $N$ for optimal graphs in Fig.~\ref{fig:graphs}. The dashed line indicates the generation cutoff at $n_G=15$, which we adopted based on early convergence.}
    \label{fig:GA_efficiency}
\end{figure}
\section{ Efficiency of the Genetic Algorithm}\label{eff:GA}
To demonstrate the efficiency of our GA in finding highly sensitive graph structures with the highest value of $D_n$, we analyze the generation index at which the maximum value of the $D_n$ is achieved for different graph sizes $N$.

Figure~\ref{fig:GA_efficiency} shows that, for most values of $N$, the optimal graphs are identified within the first few generations. This indicates that the GA converges quickly and does not require a large number of generations to locate high-performing solutions. Based on this observation, we limit our GA runs to a maximum of $n_G=15$ generations, significantly reducing computational cost without sacrificing performance.

\section{Effect of magnetic field on the graph structures}\label{effect_h}

\begin{figure}[t!]
    \centering
    \includegraphics[scale=0.45]{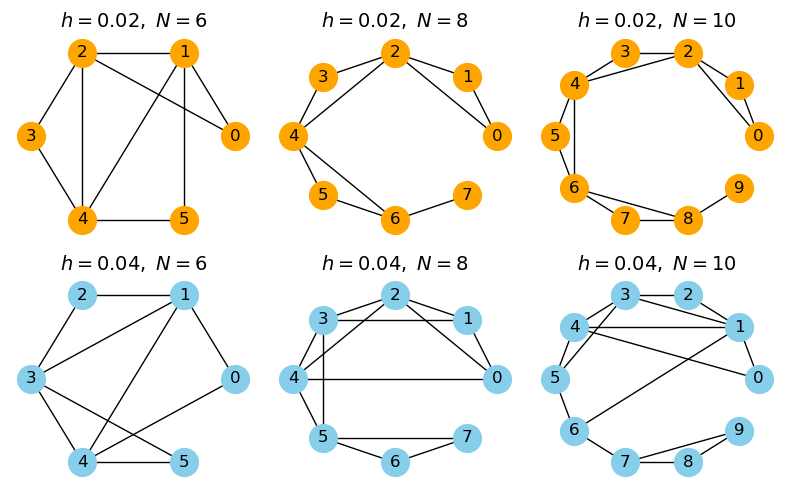}
    \caption{Graph structures optimized via GA for different values of magnetic field strength. The top row corresponds to the best-performing graphs (in terms of \( D_n \) and QFI) at $h=0.02$, while the bottom row shows the corresponding optimal graphs at higher $h=0.04$. Each column displays results for system sizes \( N = 6, 8, 10 \). The variation in edge connectivity illustrates how the strength of the magnetic field affects the optimal topology for quantum sensing tasks.}
    \label{fig:graphs_h}
\end{figure}

In this section, we present the optimized graph structures for various values of the magnetic field strength. Figure~\ref{fig:graphs_h} illustrates the structural variations of the graphs for two different magnetic field values ($h = 0.02$ and $h = 0.04$.). We observe that changes in the magnetic field strength can alter the connectivity patterns within the graphs, particularly for even system sizes. Notably, this effect becomes significant when $N \geq 6$, indicating that the magnetic field plays a nontrivial role in determining the optimal interaction network for larger systems. For odd values of $N$, the graph structures remain unchanged across variations in $h$.


\bibliography{GA}
\end{document}